\documentclass[twocolumn]{aastex61}  
\usepackage{natbib}
\usepackage{graphicx}
\usepackage{amsmath,amssymb}
\usepackage{ulem}
\usepackage{enumitem}

\newcommand{\ch}[1]{#1}
\newcommand{\cha}[1]{#1}
\newcommand{\chb}[1]{#1}

\newcommand\aastex{AAS\TeX}

\newcommand{\be}{\begin{equation}}
\newcommand{\ee}{\end{equation}}
\received{July 1, 2016}
\revised{September 27, 2016}
\accepted{December 12, 2017}
\submitjournal{ApJ}
\shorttitle{\aastex\ Turbulent heating}
\shortauthors{Montagud-Camps et al.}

\begin{document}


\title{Turbulent heating between 0.2 and 1 AU: a numerical study}

\correspondingauthor{Roland Grappin}
\email{Roland.Grappin@lpp.polytechnique.fr}

\author{Victor Montagud-Camps}
\email{Victor.Montagud-camps@lpp.polytechnique.fr}
\author{Roland Grappin}
\email{Roland.Grappin@lpp.polytechnique.fr}
\affiliation{LPP, Ecole Polytechnique, CNRS, France}
\author{Andrea Verdini}
\email{Verdini@arcetri.astro.it}
\affiliation{LPP, Ecole Polytechnique, CNRS, France
and Universit\`a di Firenze, Dipartimento di Fisica e Astronomia}

\begin{abstract}
The heating of the solar wind is a key to understand its dynamics and acceleration process. 
\cha{The observed radial decrease of proton temperature in the solar wind is} slow compared to the adiabatic prediction \cha{and it is thought to be caused by turbulent dissipation}. To generate the \cha{observed} $1/R$ decrease, the dissipation rate has to reach a specific level which varies in turn with temperature, wind speed, and \cha{heliocentric} distance.
We want to prove that MHD turbulent simulations can lead to the $1/R$ profile. We consider here the slow solar wind, characterized by a quasi-2D spectral anisotropy.
We use the EBM (expanding box model)  equations, which incorporate into 3D MHD equations the expansion due to the mean radial wind, allowing to follow the plasma evolution between 0.2 and 1 AU. We vary the initial parameters which are: Mach number, expansion parameter, plasma beta, and properties of the energy spectrum as the spectral range and slope.
Assuming turbulence starts at 0.2 AU with a Mach number equal to unity, with a 3D spectrum mainly perpendicular to the mean field, 
we find radial temperature profiles close to $1/R$ in average. This is done at the price of limiting the initial spectral extent, corresponding to the small number of modes in the inertial range available, due to the modest Reynolds number reachable with high Mach numbers.
\end{abstract}


\section{Introduction}

The radial evolution of the proton temperature in the solar wind has not yet been fully explained. From 0.2 AU 
to 1 AU, the wind expands radially and with a low-varying mean velocity. 
Observations have shown that the rate at which proton temperature decreases in this interval 
does not correspond to an adiabatic cooling of an spherically expanding \cha{flow}: instead of a temperature evolution given by $T
\propto R^{-4/3}$, a slower cooling rate is observed $T\propto R^{-\xi}$ for $\xi \in [0.8,1]$ \citep{1995JGR...100...13T}. 
The origin of this extra heating may be attributed to the development of a turbulent regime in the wind 
which generates a substantial heating
(e.g., \cite{Matthaeus:2011ip,2009PhRvL.103f1102C,2007JGRA..11207101V,2012ApJ...754...93C}) but see also \cite{2013JGRA..118.1351H} and  \cite{Scudder:2017bu} for an explanation not based on the turbulent cascade origin.

We know that in a quasi stationary incompressible turbulence, the turbulent energy (i.e., the sum of kinetic and magnetic energies) 
cascades along the inertial range without 
 being dissipated. When energy reaches the dissipation scales, it is transformed into heat. 
If the process is quasi-stationary, the heating rate 
can be obtained either from the dissipation rate \ch{(if the small-scale dissipation process is known)} or via the energy cascade 
rate.
The dissipation mechanisms take place at sub-ion scales that are not yet 
reachable by current measurements and therefore little can be said about the associated heating. 
On the contrary the cascade rate can be computed at much larger (inertial range) scales and this is the approach commonly used in analysing solar wind data to obtain the turbulent heating.

To understand if turbulent heating is responsible for the non-adiabatic decrease of temperature in the solar wind, one needs to compute the heating rate required to produce the observed temperature profile.
Such heating is obtained by exploiting an argument based on the internal energy equation (e.g., \cite{2007JGRA..11207101V}, see section~\ref{crith} below). 
In particular it shows that the existence of a radial power-law implies a direct relation between the energy cascade rate $Q$, wind velocity $U$, proton temperature $T_p$ and heliocentric distance $R$:
\be
Q = (1/2) (k_B/m) UT_p/R
\label{vasq}
\ee
the coefficient 1/2 being associated with the scaling $T_p \propto 1/R$ which we adopt here as a representative scaling.
Equation~\ref{vasq} expresses a balance between two decaying quantities: 
the energy cascade rate on the one hand, which is a turbulent quantity, and proton temperature, which is not. 

The most accurate measurements of the cascade rate are obtained by computing III-order moments of the distribution of the magnetic field \citep{1998PhRvE..57...21P}.
Such measurements have shown that the above relation eq.~\eqref{vasq} holds at 1 AU  \citep{2012ApJ...754...93C,2009ApJ...697.1119S}. 
and also at larger distances \citep{Marino_al_2008}, 
although the precise value of the cascade rate somewhat depends on hypothesis made on the 3D geometry of the angular spectra \citep{2015ApJ...804..119V}.

The existence of the $1/R$ profile between 0.3 and 1 AU suggests that the balance in eq.~\eqref{vasq} is realized also during this whole distance range,
but there is presently no measure of the cascade rate and no proof (either theoretical or numerical) that such an equilibrium is indeed achieved in the inner heliosphere.
In fact, previous attempts to verify eq.\eqref{vasq} used models of solar wind turbulence with simplified nonlinear 
couplings \citep{1988JGR....93....7T,1997SoPh..171..363T,2001JGR...106.8253S,2009JGRA..114.9103B}. 
These models actually managed to reproduce the wind 
temperature decrease, however relying on the choice of free parameters to fit their results to observations.

We aim here to examine at what conditions such a relation between turbulent heating and temperature, as well as the associated temperature profile, can be found using direct numerical simulations.
It is indeed most probable that not all turbulent conditions close to the Sun are able to generate such a close adjustment between the turbulent cascade rate and the temperature in the distance range 0.3-1 AU, so that the numerical solution to this problem will provide constraints on turbulent properties close to the Sun.

We compute directly the temperature and heating evolution vs distance, adopting successively different initial states of the plasma at the minimum heliocentric distance (here, 0.2 AU).
We use for this the MHD equations modified by expansion, as given by the expanding box model or EBM \citep{1993PhRvL..70.2190G}. 
We study in this work the case of the slow solar wind, where turbulence is mainly in a 2D geometry \citep{2005ApJ...635L.181D,2016ApJ...831..179V}.
The initial conditions found to lead to eq.~\ref{vasq} between 0.2 and 1 AU will be characterized by (i) spectral properties; (ii) global plasma properties as expansion parameter, Mach number, plasma $\beta$, and mean magnetic field angle with radial.

\section{Equations, control parameters, diagnostic tools, initial conditions}
\subsection{EBM Equations (ideal)}
We give here a short description of the expanding box model (EBM) equations \citep{1993PhRvL..70.2190G,Grappin:1996ey,Dong:2014fi} that allow to follow the turbulent evolution transported by the radial wind.
Let us denote by $\vec{U_0}=U_{0}\vec{\hat e_{r}}$ the mean wind velocity.

The wind is assumed to be radial and to have uniform speed ($U_0$=const). 
The radius R at which the box is located varies with time $\tau$ as
\be
R(\tau)=R_0 +U_0 \tau
\ee
where $R_0$ is the initial position of the box. Space, time, velocity, temperature and density are measured in the following units: 
\begin{align}
& L_0/(2\pi) \\
&t_{NL}^0 = L_0/(2\pi u_{rms}^0) \\
& u_{rms}^0 \\
& m_p (u_{rms}^0)^2/(2k_B)\\
& \rho^0 
\end{align}
where $\rho^0$ is the initial average density of the plasma, $u_{rms}^0$ is the initial rms velocity of the fluctuations, $t_{NL}^0$ is the initial nonlinear time based on the initial rms velocity, and $L_0$ is the initial size of the box \textit{perpendicular} to the radial direction.

The domain is initially elongated in the radial direction by a factor $a_x=5$, see fig.~\ref{figfig}).
Choosing the largest direction (divided by $2\pi$) perpendicular to the radial as the unit length makes sense since, as we shall see, the spectral cascade occurs preferentially perpendicularly to the mean magnetic field, \cha{and hence also perpendicularly to the radial direction (see fig.~\ref{fig6}) at small distances where the mean field is almost parallel to the radial}. 

The EBM approach relies on the idea that a simple change of Galilean frame is not sufficient to eliminate the expansion. 
After such a frame change, the plasma still expands: it is stretched in directions perpendicular to the radial.
In other words, a systematic velocity field 
perpendicular to the mean radial direction remains. To recover the usual theoretical setup where the fluctuating quantities are  homogeneous (i.e., have zero average) in the plasma volume, we need to subtract this transverse expansion. 
This is done by using  coordinates comobile with this transverse expansion:
\begin{align}
t&=\tau \\
x &= (X - U_0 \tau)/a_x \\
y &= Y/a(t) \\
z &= Z/a(t)
\end{align}
The parameter $a_x=L_x/L_y=L_x/L_z=5$ is the initial aspect ratio of the domain.
The parameter $a$ is defined as the normalized heliospheric distance: 
\be
a=R(t)/R_0=1+\epsilon t
\label{aaa}
\ee
where $\epsilon=da/dt$ is the expansion parameter defined as the initial ratio between the characteristic expansion and 
turnover times in the transverse directions (perpendicular to the radial):
\be
\epsilon=\frac{\tau_{NL}}{\tau_{exp}}=\frac{U_0/R_0}{k_{0}u_{rms}} 
\label{epsi}
\ee
with $k_0$ the minimum wavenumber in the transverse direction.
At a given distance $R(t)$, the domain has thus an aspect ratio $L_{\perp}/L_{radial} = a(t)/a_x$.

The EBM equations with dissipation terms omitted (but see below eqs.~\ref{diss1}-\ref{diss2}) read
\begin{align}
&\partial_t \rho + \nabla (\rho \vec{u}) = -2  \rho (\epsilon/a)
\label{eqro} \\
&\partial_t P + (\vec{u}.\nabla)P + \gamma P \nabla.\vec{u} = -2  \gamma P (\epsilon/a) \\
&\partial_t \vec{u} + \vec{u}.\nabla \vec{u} + \nabla (P+B^2/2)/\rho - \vec{B}.\nabla \vec{B}/\rho = - \vec{\mathbb{U}} (\epsilon/a) 
\label{eqU}\\
&\partial_t \vec{B} + \vec{u}.\nabla \vec{B} - \vec{B}.\nabla \vec{u} + \vec{B} \nabla.\vec{u} = -\vec{\mathbb{B}} (\epsilon/a)
\label{eqB} \\
&P = \rho T
\label{eqP}
\end{align}
\cha{In these equations, $\rho$ is the density, $P$ the total pressure, B the magnetic field, and u is the velocity fluctuation $u = U - U_0 \hat e_r$, $U$ being the total velocity. 
The pressure equation with $\gamma=5/3$ is the perfect gas equation. $T=T_i = T_e$ is the proton (and electron) temperature, $m_p$ being the proton mass.}

\cha{The above} equations are standard MHD equations, with, however, two modifications.
First, additional linear terms involving the constant average speed $U_0$ appear in the right-hand side:
$ \mathbb{U} = (0,u_y,u_z)$ and $ \mathbb{B} = (2B_x,B_y,B_z)$. 
Hence, depending on the component, the r.h.s. damping term differs, as is well known.
Note that in the following we will sometimes use the dimensional form of these damping terms,
namely, with $\epsilon/a(t) \rightarrow U_0/R(t)$.
Second, a new expression for the gradients is used, accounting for the increasing lateral stretching of the plasma volume with time/distance:
\be
\nabla = ((1/a_x)\partial_x,(1/a(t))\partial_y,(1/a(t))\partial_z)
\label{gradperp}
\ee
$Ox$ being along the radial, expansion acts only in the two other directions.

All fields $\rho(x,y,z)$, $u(x,y,z)$, etc.. are then considered to be periodic in all three directions of the domain comobile with the mean expansion. This allows to use a pseudo-spectral method for the spatial scheme, as in the standard Lagrangian approach.
The temporal scheme is a third-order Runge-Kutta method.
\subsection{Defining visco-resistive terms and dissipation rate}
The diffusive terms read:
\begin{align}
&\partial_t \vec{u} |_{dis} = (\mu/\tilde \rho) (\tilde \nabla^2 \vec{u} + (1/3) \tilde \nabla (\tilde \nabla \cdot \vec{u})) 
\label{diss1}
\\
&\partial_t \vec{B} |_{dis} = \eta \tilde \nabla^2 \vec{B}
\label{diss2}
\\
&\partial_t P|_{dis} = \bar \rho \kappa \tilde \Delta T
\label{diss3}
\end{align}
where $\tilde \rho$ is the density normalized by its average:
\be
\tilde \rho = \rho/\bar \rho = a^2 \rho
\ee
and $\bar \rho = 1/a^2$ is the average density of the plasma.
Note that the normalized density $\tilde \rho$ is unity in average.
\cha{Note that density enters in two different way in the dissipative terms, as the normalized density $\tilde\rho$ in the momentum equation, as the average $\bar\rho$ in the pressure equation}.
In the homogeneous (non-expanding) case, the dissipative terms given by eqs.~\ref{diss1}-\ref{diss2} are identical to the standard ones in MHD. In the expanding case, these dissipation terms have two specific differences.
First, the nabla operator components are comobile derivatives, without the anisotropic prefactors $(1/a_x, 1/a, 1/a)$: 
\be
\tilde \nabla = (\partial_x, \partial_y, \partial_z)
\label{comolap}
\ee
This allows the dissipation to be isotropic in comobile coordinates, that is, to use most efficiently the available Fourier domain.
Second, we impose that viscosity, resistivity and conductivity $\kappa$ decrease as time/distance increases (see below eq.~\ref{aaa}):
\begin{eqnarray}
\kappa = \mu =\eta= \mu_0/a 
\end{eqnarray}
This choice allows to somewhat moderate the decrease of the Reynolds number associated to the fast damping of the turbulent amplitude.

From eqs~\ref{diss1}-\ref{diss2} one derives the contribution of dissipative terms to the turbulent energy
evolution:
\be
\tfrac{d}{dt} (\rho u^2/2 + B^2/2) |_{diss} = - \bar \rho Q_\nu + \tilde \nabla \cdot F
\ee
where F is a flux which does not change the average turbulent energy.
The turbulent heating is given by the visco-resistive damping term $Q_\nu$, always positive:
\be
Q_\nu = \mu (\tilde \omega^2+ 4/3 \ (\tilde \nabla\cdot u)^{2}) + \eta \tilde J^2
\label{qnu}
\ee
where $\tilde \omega = \tilde \nabla \times u$ and $\tilde J = \tilde \nabla \times B$.

Finally, what is lost by turbulent energy is transmitted to internal energy; this reads :
\be
\partial_t P + (\vec{u}.\nabla)P + \gamma P \nabla.\vec{u} +2  \gamma P (\epsilon/a) =
\bar \rho \kappa \tilde \Delta T+ (\gamma-1) \bar \rho Q_\nu
\label{eqp}
\ee
where $ \kappa$ is the thermal conductivity.
Since we are in the following more directly interested in the temperature T, we also write down its equation:
\begin{align}
&\partial_t T + (\vec{u}.\nabla)T + (\gamma-1) T \nabla.\vec{u} +2  (\gamma-1) T (\epsilon/a) 
\nonumber \\ 
&= 
( \kappa/\tilde \rho) \tilde \Delta T+ (\gamma-1) Q_\nu/\tilde \rho
\label{eqp}
\end{align}

\subsection{Critical heating, cascade rate and parameter $M^2/\epsilon$}
\label{crith}
We here rederive eq.~\ref{vasq}, which expresses the critical heating leading to a temperature decrease as $T_p \propto 1/R$. For this we need to take the spatial average of eq.~\ref{eqp}, so eliminating the thermal conductive term. 
The other terms are checked to be negligible, e.g., using the dimensional factor $U_0/R$ instead of $\epsilon/a(t)$:
\be
\overline{\vec{u}.\nabla T} = \overline{\vec{u}.\nabla \delta T} \ll  2 \gamma \bar T U_0/R
\ee
which gives for the average temperature (the spatial average being denoted either by a bar or angular brackets):
\be
\partial_t \overline{ T} + 2 (\gamma -1)\overline{T} (\epsilon/a) = (\gamma-1) \langle Q_\nu / \tilde \rho \rangle 
\simeq (\gamma-1)\overline{Q_\nu}
\ee
where the last equality obtains assuming $\delta \rho /\rho \ll 1$.
Now we replace the temporal derivative by the radial derivative, $\partial_l \rightarrow U_0\partial_R$, and as well the damping term $\epsilon/a$ by its dimensional expression $U_0/R(t)$:
\be
U_0 \partial_R \overline{ T} + 2 (\gamma -1)\overline{T} (U_0/R) \simeq (\gamma-1)  \overline {Q_\nu}
\label{eqtbar}
\ee

We define now $Q_\alpha$ as the heating necessary to obtain a temperature profile of the form
$T = 1/R^\alpha$. Thus we replace in eq.~\ref{eqtbar} $U_0 d\bar T/dR = -\alpha U_0 \bar T/R$;
this gives:
\be
Q_\alpha = \tfrac{4-3\alpha}{2} \overline{ T} U_0/R
\label{qalfa}
\ee
In the following we will insist on the special value $\alpha=1$, which leads to what we will call the \textit{critical heating} 
$Q_c$:
\be
Q_c = Q_{1} = (1/2) \overline{ T} U_0/R
\label{qcrit}
\ee

\chb{Using simple phenomenology, we now derive a condensed formula for such a critical heating, involving basic parameters of the turbulent wind: the Mach number and the expansion parameter $\epsilon$.}
We first define the Kolmogorov cascade rate, generalized to MHD, in two ways. A first definition 
\citep{2007JGRA..11207101V} is given by
\be
Q_{K41} \simeq k (u^2+\delta B^2/\rho)^{3/2} \simeq 3 ku^3
\label{qk41a}
\ee
with $\delta B = B - \langle B \rangle$ being the magnetic field fluctuation, and both $u$ and $\delta B$ being evaluated at the Taylor's scale, thus in the inertial range. 
In eq.~\ref{qk41a}, $u^2$ and $\delta B^2/\rho$ are respectively the kinetic and magnetic energy content in the wavenumber range $[k/\sqrt 2, k \sqrt 2]$, where k lies in the inertial range.
Note that to derive the last approximate equality, we have assumed equipartition between kinetic and magnetic energy.

When analyzing our simulation results, we shall use the equivalent, but more precise definition:
\be
Q_{K41}=E(k_{\lambda})^{3/2}k_{\lambda}^{5/2} 
\label{qk41}
\ee
where $E(k_\lambda)$ is twice the 1D total (kinetic+magnetic) energy spectrum, integrated on directions perpendicular to the radial and depending on the radial wavenumber, evaluated at the Taylor's wavenumber $k_\lambda$:
The Taylor wavenumber is defined as:
\be
k_{\lambda}=\frac{ (\int_{0}^{\infty} k_{x}^2 E(k_x)dk_x )^{1/2}}{\int_{0}^{\infty}  E(k_x)dk_x } 
\label{ktay}
\ee
More precisely, $k_{\lambda}$ marks the middle of inertial zone of the reduced spectrum in the radial direction.

Now, we know from \cite{2007JGRA..11207101V} that in cold winds the Kolmogorov cascade rate $Q_{K41}$ overestimates by a factor 10 the true average dissipation rate
$\overline{ Q_\nu}$: 
\be
Q_{K41} \simeq 10 \overline{Q_\nu}
\label{vasq10}
\ee
From the latter equation, eqs~\ref{qalfa} and ~\ref{qk41a}, we rewrite the critical condition $Q_\nu=Q_\alpha$ to produce a $1/R^{\alpha}$ temperature profile as
\be
3 ku^3 \simeq 10 \  \tfrac{4-3\alpha}{2} \overline{ T} U_0/R
\label{crit1}
\ee
We want to express this condition in terms of the rms turbulent Mach number:
\be
M=u_{rms}/c_s
\label{mach}
\ee
where $c_s=(\gamma P/\rho)^{1/2}$ is the sound speed and the expansion parameter $\epsilon$ (eq.~\ref{epsi}) which also includes the rms velocity in its definition.
Assuming the inertial range begins at the largest scale, we evaluate respectively the kinetic energy content at the largest scale $u_0^2/2$ and the total energy content $u^2/2$ by integrating a Kolmogorov spectrum $\propto k^{-5/3}$ on, respectively the interval range [$k_0,2k_0$] and [$k_0,\infty$]. This leads to a ratio $(u_{rms}/u_0)^3= 4.4$, which allows to express eq.~\ref{crit1} in terms of quantities based on rms velocity amplitude, namely the Mach number M and expansion parameter $\epsilon$. One finds:
\be
M^2/\epsilon \simeq 4.4 (4-3\alpha)
\ee
and for the particular value $\alpha=1$
\be
M^2/\epsilon \simeq 4.4
\label{m2eps}
\ee
Eq.~\ref{m2eps} will guide us in selecting ``critical'' couples of initial Mach number and expansion parameter in our simulations.

\begin{figure}
\begin{center}
\includegraphics [width=0.85\linewidth]{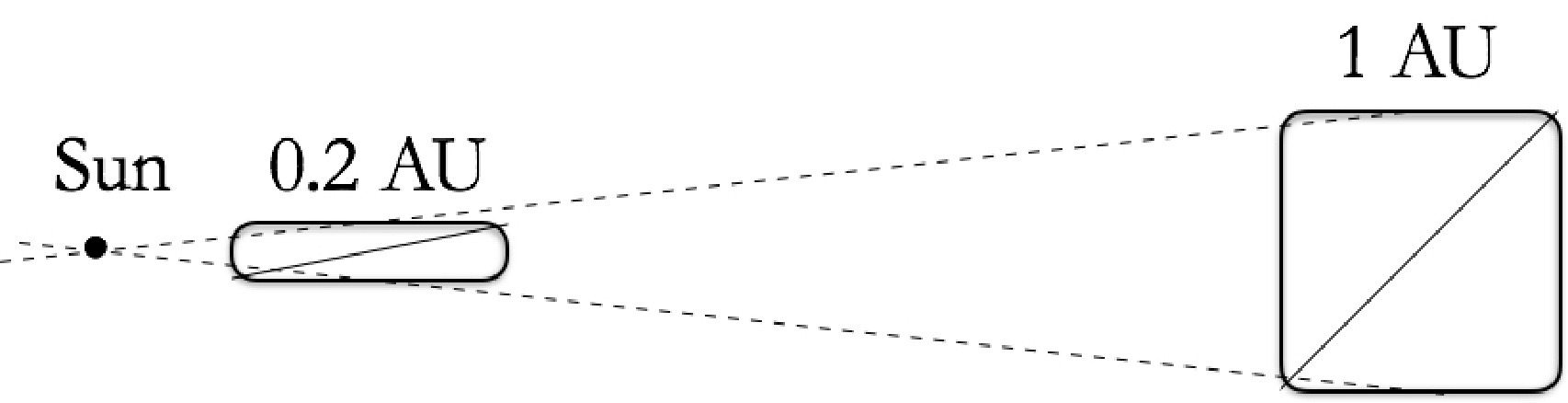}
\caption{
Initial and final domains of simulation (and plasma volume as well) in the ecliptic plane. Thin lines: direction of mean magnetic 
field. For all runs, the aspect ratio of the domain varies from 1/5 to unity. In the figure, the mean magnetic field angle with the radial 
varies varies from $\tan^{-1}(1/5) \simeq11.3^0$ to  $\tan^{-1}(1)=\pi/4$.
}
\label{figfig}
\end{center}
\end{figure}

\subsection{Turbulent energy, expansion and viscous  damping}

We define the average turbulent energy per unit mass:
\be
e = \frac{1}{2}  \langle \tilde \rho u^2 + \delta B^2/\bar{\rho} \rangle
\label{energy}
\ee
From eqs.~(\ref{eqU}-\ref{eqB}) one finds the expression of the energy damping rate due to expansion
\be
Q_{exp} =  \frac{1}{2} \langle \tilde \rho (u_y^2+u_z^2) + \delta B_x^2/\bar\rho  \rangle (2\epsilon/a) 
\ee
or, also, in dimensional terms
\be
Q_{exp} =  \frac{1}{2} \langle \tilde \rho (u_y^2+u_z^2) + \delta B_x^2/\bar\rho  \rangle (2 U_0/R) 
\ee
The two main terms leading to damping of turbulent energy are thus (i) the small scale diffusive (viscous and resistive) damping term $Q_\nu$, which is fed by the nonlinear turbulent energy cascade; (ii) the expansion damping term $Q_{exp}$. 
However, here, turbulent energy conservation is subject to further effects:
(i) compressibility breaks the turbulent energy invariance, the remaining invariant being the sum of turbulent energy and internal energy;
(ii) expansion modifies the nonlinear terms as well, thus breaking down all inviscid invariants of homogeneous MHD.
These two additional effects are gathered in a single term denoted by $Q_{NL}$. 
\chb{It is found by subtracting turbulent dissipation and linear expansion decay from the time derivative of turbulent energy}: 
\be
de/dt = - Q_\nu - Q_{exp} - Q_{NL} 
\label{damp}
\ee
In the previous equation and from now on, $Q_\nu$ (without average) will denote the spatial average $\langle Q_\nu \rangle $.
\chb{Note the very existence of a Kolmogorov-like spectral scaling $k^{-5/3}$ implies that the term $Q_{NL}$ should be subdominant compared to $Q_\nu$.}

\subsection{Numerics, initial conditions and parameters}

\begin{table}
\caption{
List of parameters for the initial conditions. 
R: name of run; $M=u_{rms}/c_s$, with $c_s$ the sound 
speed; 
$\epsilon$ is the initial expansion parameter; $M^2/\epsilon$: see section.~\ref{crith}; $B_0$ is the initial magnetic field amplitude (and Alfv\'en speed); 
$\beta$ is the ratio of thermal over magnetic pressure;
$k_{max}$ is the maximum wavenumber in directions 
perpendicular to radial (NB largest perpendicular scale corresponds to unit wavenumber); $m$ is the 1D spectral slope;
$\mu_0$ is the initial value of the diffusive parameters (viscosity, resistivity and conduction).
}
\begin{tabular}{ccccccccccc}
R & M & $\epsilon$  &$\frac{M^2}{\epsilon}$&$B_0$&$\beta$& $k_{max}$ & $m$&$\mu_0$\\
\hline
A	&$1$ 	&     0.2       &5     &2.04      &0.29	&64&$5/3$&$2.4  $ $10^{-3}$\\
B	&$1$ 	 &    0.2       &5	 &2.04      &0.29	&64&$3$   & $1.7 $ $10^{-3}$\\
C	 &$1$ 	 &  $0.2$     &5	 &2.04      &0.29	&4 &$5/3$ & $2.1 $ $10^{-3}$\\
D 	&$0.77$	 &  $0.2$     &3     &2.04      &0.49	&4 &$5/3$ & $1.8 $ $10^{-3}$\\
E	 &$1$	 & $0.2$ 	   &5	 &2.04      &0.29        &4  & 2.2   & $1.7 $ $10^{-3}$	\\
F	 &$0.77$ 	 &  $0.12$   &5     &2.04      &0.49     &4 & 2.2    & $1.8 $ $10^{-3}$	\\
G	 &$0.6$	 &  $0.14$   &2.6  &2.04	&0.8         &4 & 2.2    & $1.5 $ $10^{-3}$   	\\
H	 &$1$ 	 &  $0.4$	   &2.5  &2.04      &0.29     &4 & 2.2    & $1.3 $ $10^{-3}$	\\
K 	& $0.77$ 	 &  $0.2$     &3	 &2.04      &0.49	&4 & 2.2    & $1.8$ $10^{-3}$ \\
M	&$0.77$ 	 &  $0.2$     &3     &0.86	&2.75	&4 & 2.2    & $1.8$ $10^{-3}$ \\
N	&$0.77$ 	 &  $0.2$     &3     &1.17	&1.48	&4 & 2.2    & $1.8$ $10^{-3}$ \\
Z	&$0.3$ 	 &  $0.2$     &0.45&2.04	&0.29	&64&$5/3$& $1.5$ $10^{-4}$ \\
\end{tabular}
\label{table1}
\end{table}

All simulations are computed in a numerical box with resolution $N_x=N_y=N_z=512$. 
We  start at 0.2 AU with a numerical box elongated with an aspect ratio $a_x=5$ along the radial, of dimensions $L_x=5 \times 
L_y=5 \times L_z=5\times 2\pi$. 
The domain is then stretched by expansion in directions perpendicular to the radial so that at 1 AU the domain becomes a cube, 
and the mean field rotates accordingly (fig.~\ref{figfig}). 

We set up energy equipartition:
$u_{rms}=b_{rms}=1$ and about zero correlation between magnetic and velocity fluctuations (zero cross-helicity).
Energy isocontours are spatially anisotropic, having the 
same aspect ratio as the numerical box. 
As the mean magnetic field makes in all runs a small angle $\theta$ with the radial (between $11^0$ and $20^0$), the initial spectrum has isocontours not far from perpendicular to the radial direction with aspect ratio equal to 5 (see fig.~\ref{fig7}a).
This corresponds to the so-called "2D" configuration characteristic of slow winds 
\citep{2005ApJ...635L.181D,2016ApJ...831..179V}.

The parameters of the simulations are listed in Table~\ref{table1} which lists all runs described here.
The main parameters are the Mach number and expansion parameter $\epsilon$.
In Table~\ref{table1}, $\epsilon$ goes from 0.12 to 0.4, $M$ from 0.3 to 1, $M^2/\epsilon$ from 0.45 to 5.
Varying $\epsilon$ by a factor 2 as done here corresponds to vary the initial turnover time of the largest scale by such a factor, so the wavenumber by a factor $2^{2/3}$, adopting a $k^{-5/3}$ scaling for the initial spectrum. Also, the simulation duration is longer when $\epsilon$ is smaller, as the transport distance (from 0.2 to 1 AU) is fixed.

Each simulation provides a test window of wavenumbers (reduced to a little more than two decades)
on the much larger solar wind inertial range. When choosing the couple $(M, \epsilon)$ for a given run, we choose
a wind regime, and at the same time we place our simulation range on the solar wind wavenumber range.
Typical values, as taken in the four runs A, B, C, E are M=1 and $\epsilon=0.2$.
Such values are not far from values found in cold winds in Helios data at the scale of several hours \citep{1991AnGeo...9..416G}.

As seen previously in section~\ref{crith}, the expression $M^2/\epsilon$ was found to be approximately 4.4 in 
the inertial range at 1 AU by \cite{2007JGRA..11207101V} for cold/slow winds.
This is in the middle of the range of values listed in table~\ref{table1}.

Other parameters appearing in Table~\ref{table1} are: the plasma beta, $\beta=P_{th}/P_{B}= (2/\gamma)(c_s/v_a)^2$ which 
is varied from 0.29 to 1.48, the mean field $B_0$ which varies  between 0.86 and 2.04, which means, since $b_{rms}=1$, that 
$b_{rms}/B_0$ varies from 1.2 to 0.5.
The initial viscosity $\mu_0$ is about the same in all calculations except for run Z: the origin of this exception will be discussed in Section~\ref{spectral}.
The parameters $k_{max}$ and $m$ are respectively the extent of the initial spectrum and its 1D 
spectral slope.

\section{Results}

\subsection{Run A: \cha{extended initial spectrum}}
We start with the case of run A, with $M=1$, $\epsilon=0.2$, thus $M^2/\epsilon=5$.
The initial spectral extent in the direction perpendicular to the radial is $k_{max}=64$, spectral slope is $m=5/3$.

\begin{figure}
\begin{center}
\includegraphics [width=0.45\linewidth]{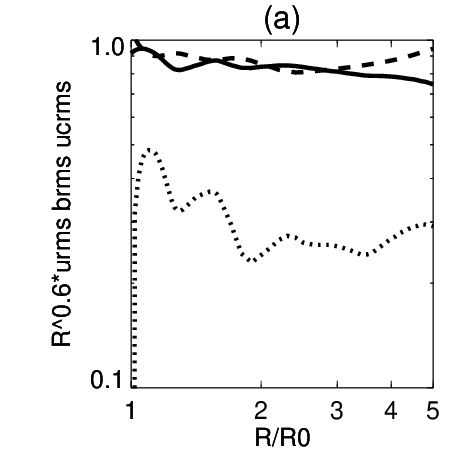}
\includegraphics [width=0.45\linewidth]{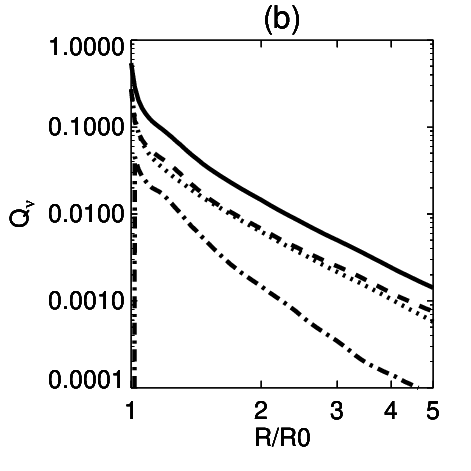}
\includegraphics [width=0.45\linewidth]{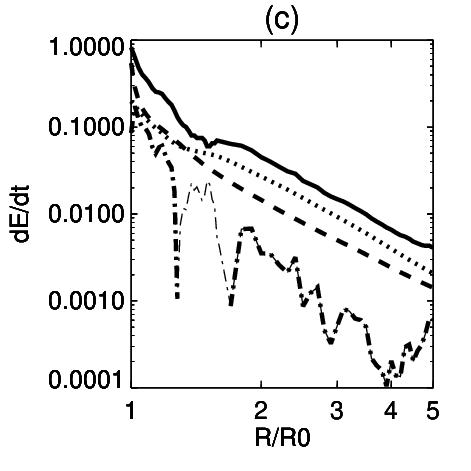}
\includegraphics [width=0.45\linewidth]{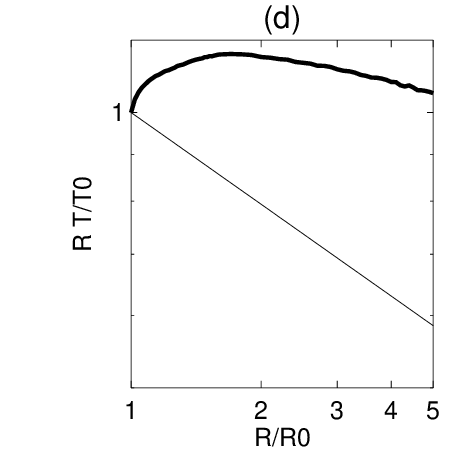}
\caption{
Run A. Evolution of basic quantities vs heliocentric distance $R/R_0$.
 (a) Velocity amplitude $u_{rms}$ (solid line), compressible velocity $u^c_{rms}$ (dotted) and magnetic field fluctuation $b_{rms}/\bar \rho^{1/2}$ (dashed).
 (b) Visco-resistive dissipation $Q_\nu$ (solid line) decomposed as the sum of $\mu |\tilde \nabla \times u|^2$ (dotted), 
$\eta \tilde j^2$ (dashed) and $\frac{4}{3}\nu |\tilde \nabla .u|^2$ (dotted-dashed).  
 (c) Dissipation rates per unit mass: $dE/dt$ (solid line), expansion-driven damping $Q_{exp}$ (dotted), visco-resistive dissipation 
 $Q_\nu$ (dashed), nonlinear loss during cascade $Q_{NL}$ (dotted-dashed, thick when increasing the decay rate, thin when decreasing the decay rate).
 (d) Temperature compensated by $1/R$ decrease.
Distance is normalized by the initial distance $R_0$ = 0.2 AU.
}
\label{fig10}
\end{center}
\end{figure}

Fig.~\ref{fig10} shows the evolution of several quantities vs heliocentric distance. 
Panel (a) shows the evolution of the rms turbulent quantities: $u_{rms}$, $\delta B_{rms}/\bar{\rho}^{1/2}$ and $u^c_{rms}$
where $u^c$ is the compressible part of the velocity field. The quantities shown are multiplied by $(R/R_0)^{0.6}$.
One sees that the \cha{compensated profiles of velocity and magnetic field (in velocity units)} are close to a plateau, \cha{indicating a decay close to $1/R^{0.6}$}, somewhat faster than the WKB prediction
$u_{rms} \simeq b_{rms}/\bar \rho^{1/2} \simeq 1/R^{1/2}$.
The compressive rms velocity amplitude, initially zero, reaches rapidly about half that of the total rms velocity and then becomes closer to 1/3 of it.
Panel (b) shows the average turbulent turbulent dissipation of total energy (solid line), solenoidal kinetic ($\mu \tilde \omega^2$, dotted line), magnetic ($\eta \tilde J^2$, dashed line) and compressible energy ($4/3 \mu (\tilde \nabla \cdot u)^2$, dotted-dashed line). 
While the kinetic and magnetic dissipation are comparable as expected, the compressible dissipation decreases rapidly
and becomes 1/10 of the total dissipation at the end.

Panel (c) of fig.~\ref{fig10} shows the total energy (eq.~\ref{energy}) decay rate $|de/dt|$ (solid line) and its components: the expansion decay rate $Q_{exp}$ (dotted), the turbulent decay rate again $Q_\nu$ (dashed) and finally the residual term $Q_{NL}$ (dotted-dashed). 
\chb{One can distinguish two phases: (a) a short initial transient, during which the turbulent dissipation dominates the expansion decay, in agreement with the small value of the expansion parameter $\epsilon=0.2$, and a very large residual term comparable to the 
turbulent dissipation: $Q_{exp} < Q_\nu \simeq Q_{NL} $; 
(b) the rest of the evolution during which turbulent decay is smaller than the expansion decay, and the residual decay is the smallest:
$Q_{NL} < Q_\nu < Q_{exp}$.
These two points will be clarified in the discussion.}
Remark also that the sign of $Q_{NL}$ varies: it is an energy loss (thus increasing $|de/dt|$), denoted by a thick line, but during the beginning phase ($1 \le R/R_0 \le 2$, thin line), it is an energy gain, thus decreasing $|de/dt|$.

Finally, panel (d) gives the resulting temperature curve, compensated by a $1/R$ law. 
One sees that a power-law regime appears for $R \ge 0.5$ AU, with an index in between $4/3$ and $1$.
The turbulent energy reservoir is clearly used in two phases: (i) an early phase with rapid and strong dissipation which 
almost stops the plasma cooling; (ii) a long-lasting phase with reduced dissipation which delays only mildly the cooling of the plasma.

\subsection{Varying spectral extent}
Is the set of initial conditions made for run A the most efficient in terms of resulting temperature curve, or can we achieve a resulting curve closer to the observed $1/R$ decrease?

As we will see later in the discussion, an important characteristic of run A is the exaggerated importance of small scales, compared to the one found in the quasi-stationary turbulent state.
This is at the origin of the excessive heating occurring in the early phase of the run.
To reduce energy of small scales, we now change one or two of the following parameters:
(i) the initial spectral slope $m$; (ii) the initial power-law extent, as measured by $k_{max}$.

\begin{figure}
\begin{center}
\includegraphics [width=0.48\linewidth]{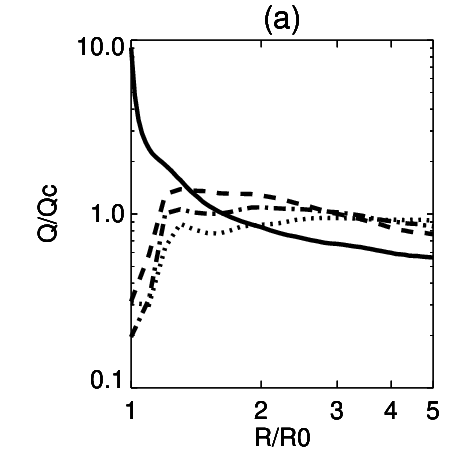}
\includegraphics [width=0.48\linewidth]{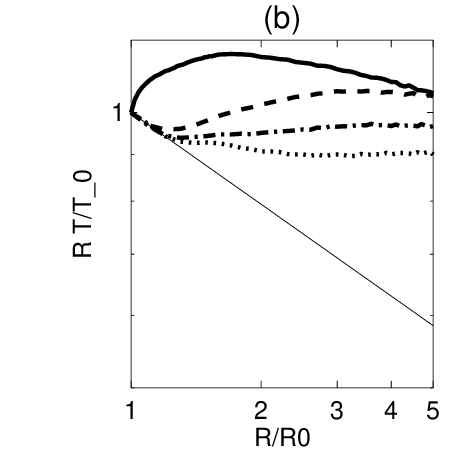}
\caption{
Heating ratio $Q_{\nu}/Q_c$ and temperature profiles, all with $M=1,\epsilon=0.2$ but varying small-scale initial excitation.
Runs A (thick solid line), B (dotted), C (dashed), E (dashed-dotted).
(a) Heating ratio $ Q_{\nu}/Q_{c}$ vs heliospheric distance $R$. 
(b) Average temperature (normalized by its initial value) compensated by $R_0/R$. 
Distance is normalized by the initial distance $R_0=0.2$ AU. 
The thin solid straight line in panel (b) corresponds to $T/T_0=(R/R_0)^{-4/3}$.
}
\label{fig1}
\end{center}
\end{figure}

We compare in fig.~\ref{fig1} the four runs A (solid), B (dotted), C (dashed), E (dotted-dashed), with the following values of $m$ and $k_{max}$:
$m=5/3$ and $k_{max}=64$ for run A, $m=3$, and $k_{max}=64$ for run B, 
$m=5/3$ and $k_{max}=4$ for run C, and $m=2.2$ and $k_{max}=4$ for run E.
Panel (a) shows the critical heating ratio $Q_\nu/Q_c$ which in principle reveals how close we are from critical heating; 
panel (b) shows the temperature curve compensated for $1/R$ decrease.

Run A shows an initial large overheating phase with initially $Q_\nu/Q_c \simeq 10$, followed by insufficient heating 
$Q_\nu/Q_c < 1 $ for $R > 0.3 AU$ (panel a, solid line).
This explains the different phases of temperature evolution already considered earlier: 
a large part of the turbulent energy is lost during the first phase, and so in the second phase the remaining 
turbulent energy is too small to heat substantially the plasma, leading to a temperature decrease 
in between adiabatic and $1/R$ (panel b, solid line).

Due to the reduced importance of small scales, runs B, C, and E show a different behavior.
Initially, for $R\lesssim1.2$, all of them show comparable critical heating ratio \cha{(panel a)}, too small to lead to \cha{an} observable heating. \cha{This phase corresponds to a quasi-adiabatic decrease of temperature (panel b)}. 
This is followed by a quasi-stationary regime \cha{($R\gtrsim1.2$)}, in which the heating is close to critical \cha{and leads} to a common temperature decrease, with all three temperature curves showing a quasi-plateau, thus close to a $1/R$ decrease (panel b).

Decreasing the importance of small scales in the initial spectrum thus succeeds in suppressing the too large energy loss of the first phase.
Note that run E shows the profile closest to $1/R$ during the whole non-adiabatic phase.
In the following runs, we thus fix the spectral parameters as in run E: spectral slope $2.2$, and a short spectral extent with 
$k_{max}=4$.

\subsection{Mach number, $\epsilon$, and $M^2/\epsilon$}

\begin{figure}
\begin{center}
\includegraphics [width=0.48\linewidth]{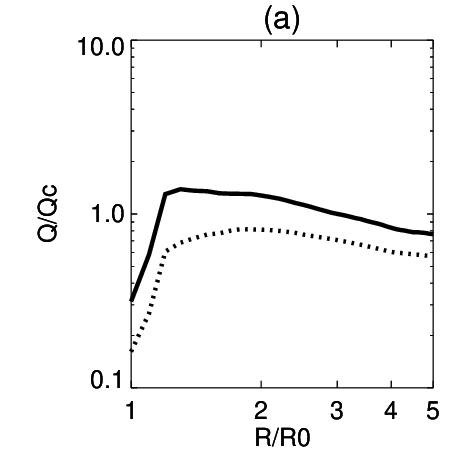}
\includegraphics [width=0.48\linewidth]{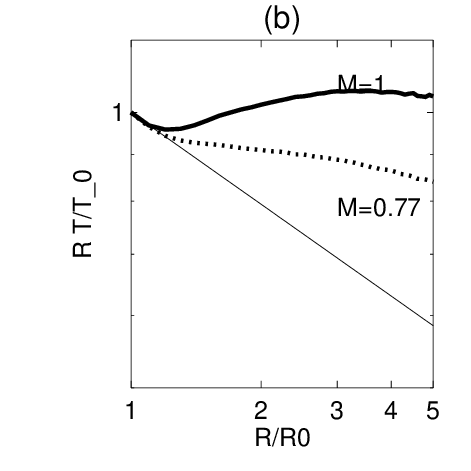}
\caption{Heating ratio $Q_{\nu}/Q_c$ and temperature profiles, with same $\epsilon=0.2$ but varying initial Mach number.
Runs C (M=1, solid thick line) and D (M=0.77, dotted).
Same caption as in fig.~\ref{fig1}.
}
\label{fig3}
\end{center}
\end{figure}

Decreasing the initial Mach number intuitively decreases the turbulent energy reservoir compared to the internal energy, so it 
should also decrease the heating ratio $Q_{\nu}/Q_c$. 
In order to check this conjecture, we compare two runs, C and D, 
both with $k_{max}=4$, $\epsilon=0.2$, and respectively $M=1$ and $M=0.77$.
Fig.~\ref{fig3}a shows that our conjecture for the heating ratio is correct.
As a consequence, the average radial slope of the temperature profile changes substantially (Fig.~\ref{fig3}b).
Note however that the temperature curve shows a break and decreases at a faster rate in the end.

\begin{figure}
\begin{center}
\includegraphics [width=0.48\linewidth]{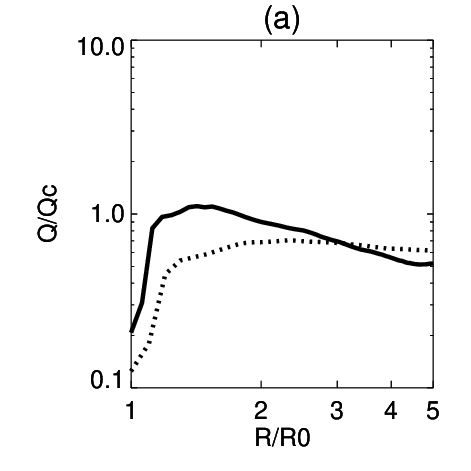}
\includegraphics [width=0.48\linewidth]{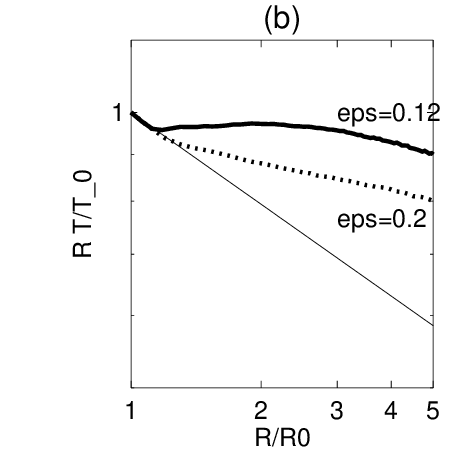}
\caption{Heating ratio $Q_{\nu}/Q_c$ and temperature profiles, with same $M=0.77$ but varying expansion parameter.
Runs F ($\epsilon=0.12$, solid thick line) and K ($\epsilon=0.2$, dotted);
same caption as in fig.~\ref{fig1}.
}
\label{fig4}
\end{center}
\end{figure}

The expansion parameter $\epsilon$ measures the expansion rate normalized by the nonlinear shearing rate.
Intuitively again, a low expansion parameter should favor heating to the detriment of cooling.
In order to check this second conjecture, we compare two values of the expansion parameter: 
$\epsilon=0.12$ (run F) and $\epsilon=0.2$ (run K), with $M=0.77$ in both cases.
The result is as expected (fig.~\ref{fig4}), i.e., the heating is larger for run with lower $\epsilon$,
during the first part of the transport, for $R < 0.6$ AU ($R/R_0 < 3$).
However, the reverse is true for the second half of the travel.
\chb{This happens because for smaller values of $\epsilon$ (run F, solid line) the same travel distance corresponds to a larger number of nonlinear times (i.e., larger ``age'', see \cite{1991AnGeo...9..416G}), which may easily result in a too fast decrease of the energy reservoir, and thus of the heating rate.}

Last, we test the parameter $M^2/\epsilon$ as a possible control parameter for the heating and temperature profile (see 
Section~\ref{crith}).
We choose two pairs of runs: E and F have the largest parameter value $M^2/\epsilon=5$ and runs G and H have the lowest one: 
$M^2/\epsilon \simeq 2.5$ (see table~\ref{table1}).

\begin{figure}
\begin{center}
\includegraphics [width=0.48\linewidth]{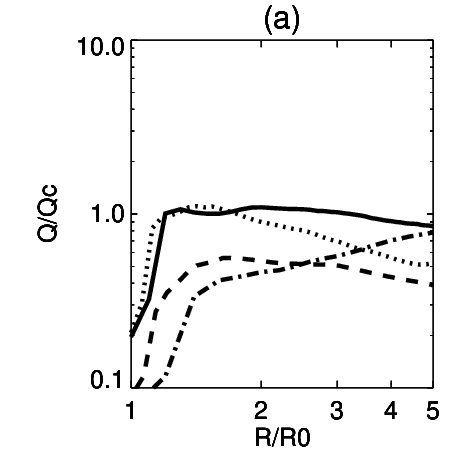}
\includegraphics [width=0.48\linewidth]{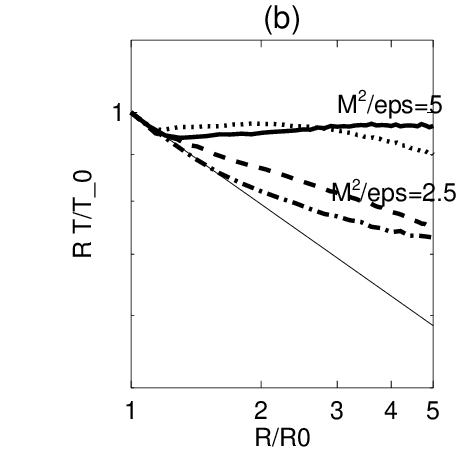}
\caption{Heating ratio $Q_{\nu}/Q_c$ and temperature profiles, with varying $M^2/\epsilon$.
Runs E (solid thick line), F (dotted), G (dashed), H (dotted-
dashed).
Same caption as in fig.~\ref{fig1}.
}
\label{fig2}
\end{center}
\end{figure}

Fig.~\ref{fig2}a shows that the ordering of critical heating by $M^2/\epsilon$ is approximately verified for $R<0.5$ AU, but not at larger distances.
Nevertheless, the temperature curves (fig.~\ref{fig2}b) appear to be gathered in two groups according to the parameter value.

\subsection{Varying the plasma $\beta$ and angle $\theta_{VB}$}
\begin{figure}
\begin{center}
\includegraphics [width=0.48\linewidth]{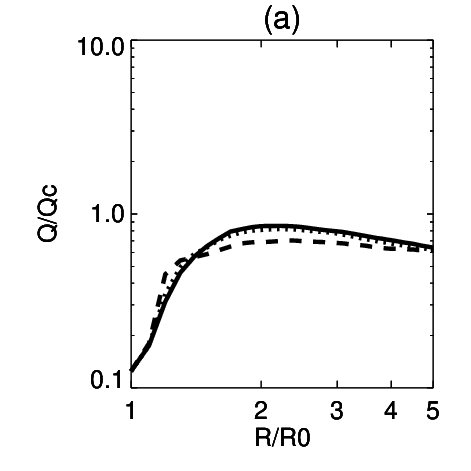}
\includegraphics [width=0.48\linewidth]{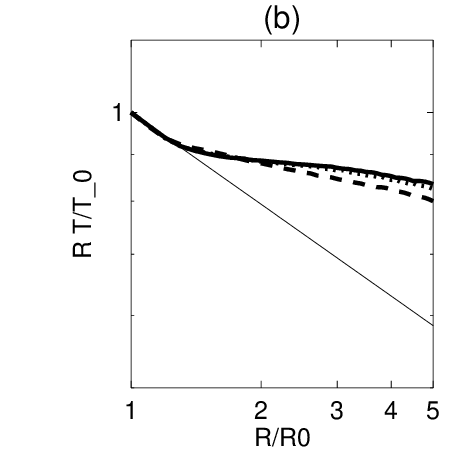}
\caption{
Heating ratio $Q_{\nu}/Q_c$ and temperature profiles, with $M=0.77$, $\epsilon=0.2$, but varying $\beta$.
Runs M (solid thick line, $\beta=$2.75), N (dotted, $\beta=$1.48) and K (dashed, $\beta=$0.49).
Same caption as in fig.~\ref{fig1}.
}
\label{fig5}
\end{center}
\end{figure}

Due to variations of the initial temperature, with the mean initial magnetic field 
remaining constant, the $\beta$ of the different runs considered up to now (runs A to K) 
has been varied up to now in the interval $0.3 \le \beta \le 0.8$ (table~\ref{table1}).
Since in the slow solar wind, the $\beta$ of the plasma can be larger than unity, we consider now runs with larger values: $\beta=2.75, 1.48$ (runs M, N) and compare with run K with $\beta=0.49$. Mach number is 0.77 in the three runs.
The overall effect of the $\beta$ variation appears to be small (fig.~\ref{fig5}), with however a slight advantage (stronger heating) to the two runs with larger $\beta$ (runs M and N, thick and dotted lines, with quasi-superposed curves).

A last parameter is of interest: it is well known that, although the angle $\theta_{VB}$ between the magnetic field and the radial 
direction is on average $45^0$ at 1 AU (which corresponds to 
$\theta=11^0$ at 0.2 AU), its distribution actually varies widely around the average.
To test the effect of small angle variation, we considered doubling the initial angle $\theta_{VB}^0$.
We found that when passing from $\theta_{VB}^0=11^0$ up to $\theta_{VB}^0=20^0$, the critical heating ratio as well as the 
temperature profile show no variation at all (not shown).

\section{Discussion}

\subsection{Summary}

Our numerical results support the possibility that MHD turbulence can drive a proton temperature profile decreasing significantly more slowly than the adiabatic prediction, in the distance range $0.2 < R < 1$ AU.
We started with a spectrum initially having a 2D configuration, corresponding in principle to the slow wind regime as observed by 
\cite{2005ApJ...635L.181D}.
This led, with an rms Mach number close to unity and expansion parameter $\epsilon=0.2$, to a temperature profile significantly steeper than observed, however when considering a strong reduction of the initial spectral inertial range, we obtained a temperature profile close in average to a $1/R$ law, thus not far from the average $1/R^{0.9}$ profile measured by \cite{1995JGR...100...13T}.

We found that the parameters \cha{regulating the heating rate} are the rms Mach number and the expansion parameter $\epsilon$, combined as $M^2/\epsilon$, \cha{while} other parameters as plasma $\beta$, angle $\theta_{VB}$ between the mean field and the radial (for small initial values) have a minor effect.

\subsection{Spectral properties vs Mach number}
\label{spectral}

\begin{figure}
\begin{center}
\includegraphics [width=0.48\linewidth]{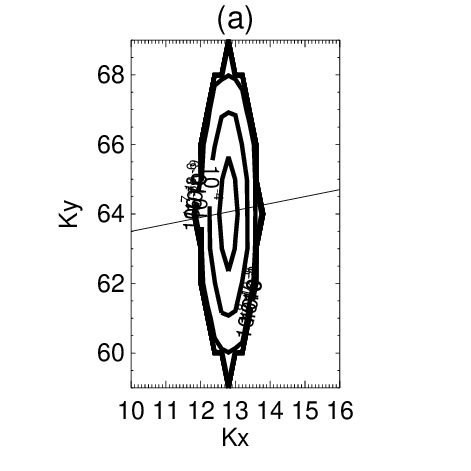}
\includegraphics [width=0.48\linewidth]{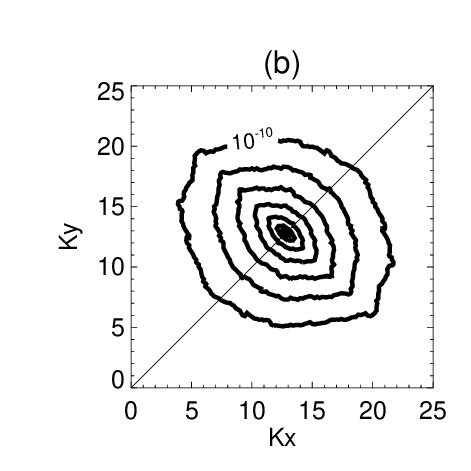}
\caption{
Run E: rotation of the energy spectrum (kinetic + magnetic) with the mean magnetic field. 
Angular energy spectrum E($k_x$,$k_y$) in the plane $k_z$=0. 
(a) At R=0.2 AU;
(b) at R = 1 AU. 
The mean magnetic field direction is represented by a straight line in each panel.
}
\label{fig6}
\end{center}
\end{figure}

\begin{figure}
\begin{center}
\includegraphics [width=0.48\linewidth]{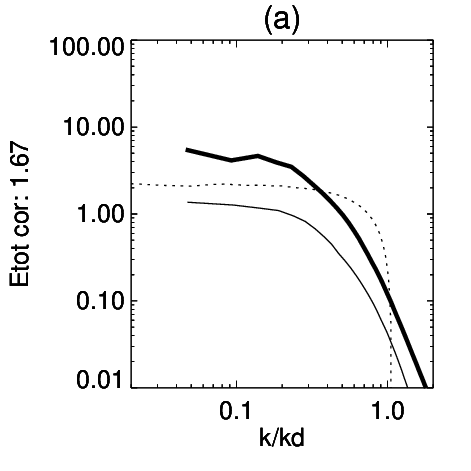}
\includegraphics [width=0.48\linewidth]{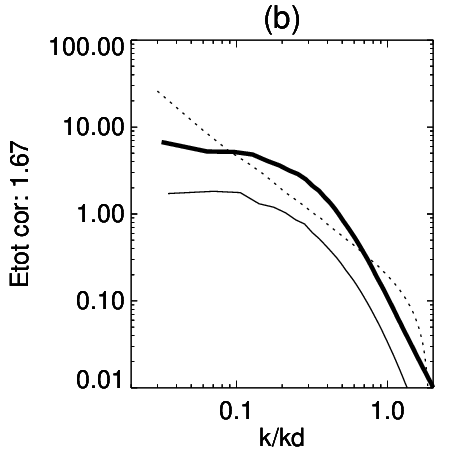}
\includegraphics [width=0.48\linewidth]{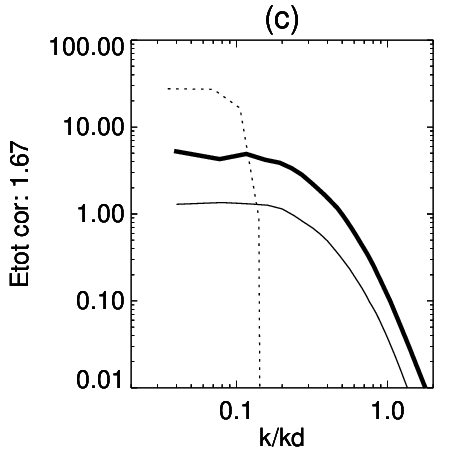}
\includegraphics [width=0.48\linewidth]{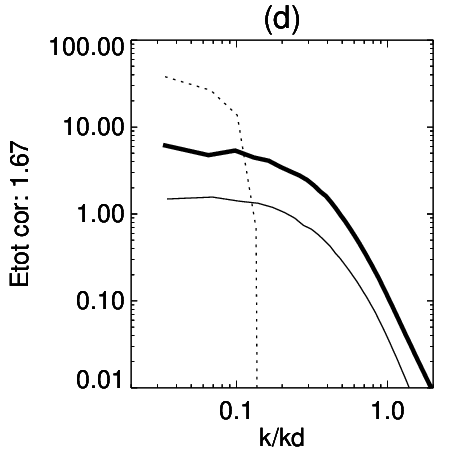}
\caption{
Runs A, B, C, E: evolution of the 1D energy spectrum (kinetic + magnetic), compensated by the Kolmogorov scaling $k^{-5/3}$.
Abscissa: wavenumber normalized by the Kolmogorov wavenumber $k_d=(Q_\nu/\mu^3)^{1/4}$.
Dotted line: $R=R_0=0.2 AU$; Solid lines: $R=1 AU$; thick solid line: radial 1D spectrum; thin solid line: 1D spectrum 
perpendicular to radial ($\hat{z}$ direction).
}
\label{fig7}
\end{center}
\end{figure}

To understand the \chb{necessity of reducing} the initial small-scale energy content,
we examine here the spectral evolution.

We first examine the spectral anisotropy. Figure~\ref{fig6} shows the case of run E; it is representative of the other runs, that all show comparable evolutions.
The 2D spectra shown are cuts though the plane $k_z=0$ of the 3D spectrum for total energy $u^2+\delta B^2/\rho$.
As explained in section 2, the initial energy isocontours (panel a, R=0.2 AU) are quasi-perpendicular to the mean field direction (thin line) which is close to the radial direction. 
At 1 AU however (panel b), the mean magnetic field has an angle of $\pi/4$ with the radial, and the main symmetry axis of the isocontours is now quasi-perpendicular to this mean field.
In other words, the cascade is not only perpendicular initially to the mean field, it remains so during transport,
following the rotation of the mean field.
This corresponds nicely to the so-called 2D spectrum dominant in slow wind \citep{2005ApJ...635L.181D}
which has first been found numerically in \citep{2016ApJ...831..179V} to be one of the two robust attractors in the wind.

We now consider the 1D reduced total energy spectra at 0.2 and 1 AU for runs A, B, C and E, shown in fig.~\ref{fig7}.
Initial spectra have dotted lines, and final ones have solid lines (1 AU). 
Final spectra depend on either radial (thick solid line) or perpendicular wavevectors (thin solid line).
The wavenumber is normalized by the Kolmogorov dissipation wavenumber $k_d=(Q_\nu/\mu^3)^{1/4}$ where the dissipation rate $Q_{\nu}$ is defined in eq.~\ref{qnu}. 
Each final spectrum is then obtained by averaging in the distance interval $0.6 \le R \le 1$ the spectra so normalized.

The four final spectra are all comparable (either along radial or transverse direction), showing a very reduced spectral extent of about half a decade with a slope $m \gtrsim 5/3$.
\begin{figure}
\begin{center}
\includegraphics [width=0.7\linewidth]{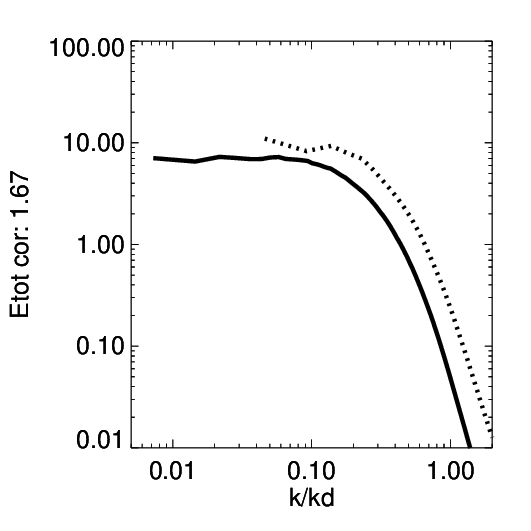}
\caption{
Runs Z (thick solid line) with $M=0.3$, A (dotted line) with $M=1$: total 1D energy spectrum averaged during the last phase of 
transport, compensated by the $k^{-5/3}$ scaling;
abscissa: radial wavenumber normalized by the Kolmogorov wavenumber. 
}
\label{fig9}
\end{center}
\end{figure}

The origin of the small extent of the inertial range in the final spectra actually lies in the high Mach number (M=1) adopted in these runs. 
A high Mach number lead to large intermittent variations of density associated to shocks, thus requiring large 
viscosities to prevent the occurrence of unresolved gradients in high density regions.
As a matter of comparison, when dealing with run Z (M=0.33), we could use a viscosity \cha{\textit{ten} times smaller than that used for runs with $M=1$ (see table~\ref{table1})}.
Due to its much lower viscosity, run Z follows a $k^{-5/3}$ scaling on more than one decade, thus significantly larger than for run A (fig.~\ref{fig9}).

The short spectral extent seen previously for runs with $M=1$ thus results from the necessity to 
increase the viscosity with such large Mach numbers to prevent a catastrophic (unphysical) evolution of the run at a given numerical resolution.
This also explains why it is necessary to start with a small spectral extent: otherwise, one obtains during a transient phase 
an excessive heating produced by the artificial initial excess of energy at \cha{visco-resistive scales (see the evolution of $Q_{\nu}/Q_c$ for run A in fig.~\ref{fig1}a).}

\subsection{Dissipation rate and Kolmogorov rate}
\begin{figure}
\begin{center}
\includegraphics [width=0.49\linewidth]{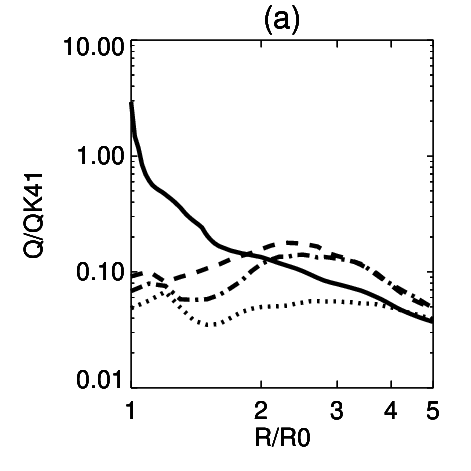}
\includegraphics [width=0.49\linewidth]{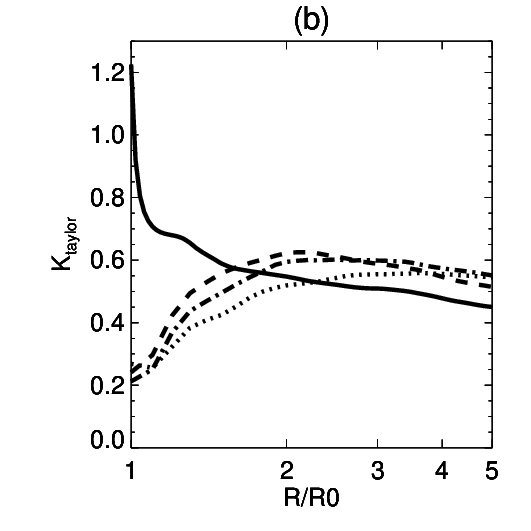}
\includegraphics [width=0.49\linewidth]{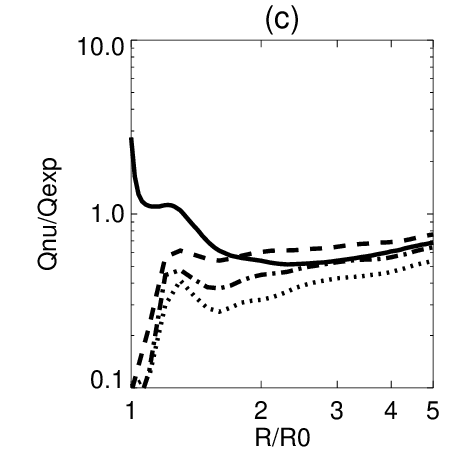}
\includegraphics [width=0.49\linewidth]{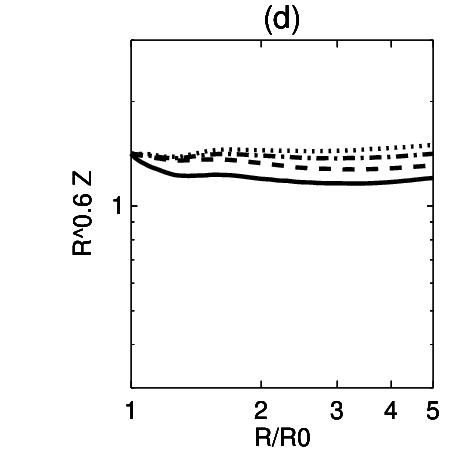}
\caption{Checking Vasquez law and consequences.
(a) Ratio $Q_\nu/Q_{K41}$ between visco-resistive dissipation and Kolmogorov's energy cascade rate 
(eq.~\ref{qk41}). 
(b) Radial Taylor wavenumber;
(c) Ratio $Q_\nu/Q_{exp}$ between visco-resistive dissipation and expansion decay rate; 
(d) Turbulent rms amplitude $Z=(u^2+\delta B^2/\overline \rho)^{1/2}$ evolution compensated by $1/R^{0.6}$.
Runs A (solid line), B (dotted line), C (dashed line), E (dotted-dashed line).
}
\label{fig12}
\end{center}
\end{figure}

The parameter $M^2/\epsilon$ (eq.~\ref{m2eps}), derived in Section~\ref{crith}, has been used to specify conditions allowing to approach critical heating. In particular, in runs B, C, E the value $M^2/\epsilon=5$, (not too far from the nominal value of $4.4$ derived in eq.~\ref{m2eps}), allowed us to obtain temperature profiles with power-law close to $1/R$.

The argument used in Section~\ref{crith} leading to this prescription relies on the assumption that the ratio 
$R_V = Q_\nu/ Q_{K41}$ is $ \simeq 0.1$. This very low value of $R_V$ (corresponding to an effective very high Kolmogorov constant), has been found to hold in cold winds by \cite{2007JGRA..11207101V}.
It would be satisfying to check whether or not the runs studied in this paper do show the same low value of $R_V$.

Fig.~\ref{fig12}a gives $R_V$ vs distance in the four runs A, B, C, E.
The curve varies wildly for run A, $R_V$ passing from larger than 1 to lower than 0.1, while runs B, C, E all show a ratio clustering around the value 0.1 on the whole distance range.
Panel (b) shows the evolution of the Taylor wavenumber (eq.~\ref{ktay}) for the four runs, summarizing the spectral width evolution for the different runs.
By comparing panels \cha{(b) and (a)}, one sees again that the initial oversized spectral width of run A leads to an anomalously large value of $R_V$.
In contrast, runs B, C, E which have a spectral width adapted to their viscosity show $R_V$-values clustering around 0.1.

The fact that, in our simulations B, C, E leading to a critical heating, we find a value for $Q_\nu/Q_{K41}$ close to observed indicates that our numerical setup is, in spite of the previous remarks on the limited spectral range, leads to turbulent properties close to those of the actual solar wind turbulence.

This small value of $R_V$ means that the characteristic turbulent decay time is about ten times longer than the plain non-linear time $t_{NL} = 1/ku$.
This allows us to interpret the relative importance in our runs of the expansion decay rate and turbulent decay rate.
Indeed, with an expansion parameter $\epsilon=0.2$ as in runs A, B, C, E, one expects a priori an expansion decay rate $Q_{exp}$  smaller than the turbulent dissipation rate, since $\epsilon$ is the expansion rate normalized by the inverse of the turnover time at large scales (eq.~\ref{epsi}).
However, due to $R_V$ being 0.1, the effective turbulent decay time scale is 10 \cha{times longer} than the nonlinear turnover time. 
The effective expansion decay rate is thus finally not smaller, but larger than the turbulent dissipation rate, in spite of $\epsilon=0.2$.
This is true in the whole distance range [0.2,1] AU for runs B, C, E, and as well for run A, except during the early transient where $R_V$ is close to unity.

A corollary to the overall dominance of expansion damping in runs A, B, C, E, should be that the turbulent fluctuation amplitude decays close to the wkb prediction which is for Alfv\'en waves 
\be
Z=(u^2+\delta B^2/\overline \rho)^{1/2} \simeq 1/R^{1/2}
\ee
This is indeed the case: Fig.~\ref{fig12}d shows that the turbulent amplitude in runs A, B, C, E decays as
$Z \simeq 1/R^{0.6}$, thus close to the wkb prediction.

\subsection{Loss of energy conservation during cascade}
\begin{figure}
\begin{center}
\includegraphics [width=0.49\linewidth]{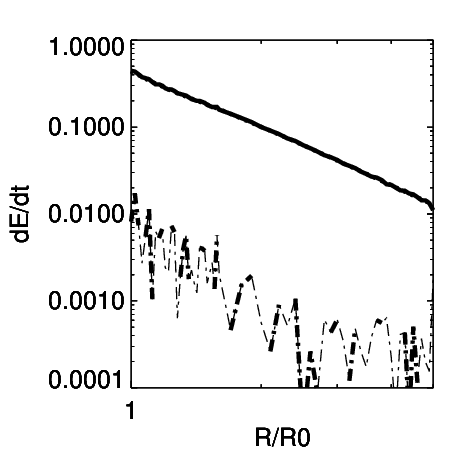}
\caption{Run A. Deviation from total (turbulent + internal) energy conservation during cascade.
Solid line: time derivative of total energy (turbulent + internal, eq.~\ref{damp2};
dotted-dashed line (thick when increasing the decay rate, thin when decreasing the decay rate): residual term $Q_{NL'}$ measuring the deviation from total energy conservation during the cascade apart from linear effects.
}
\label{fig13}
\end{center}
\end{figure}

\chb{The deviation from turbulent energy conservation during cascade has been measured by the residual term $Q_{NL}$ defined as the difference between the total turbulent energy decay and the sum of turbulent dissipation $Q_\nu$ and linear expansion decay $Q_{exp}$ (eq.~\ref{damp}, run A, fig.~\ref{fig10}c).}
Other runs show that with a fixed Mach number, the residual term $Q_{NL}$ is proportional to the expansion parameter $\epsilon$.
In order to eliminate the contribution of compressibility and so to determine without ambiguity the contribution of expansion alone, we consider the deviation of total energy conservation instead of just turbulent energy. We denote the new residual term by $Q_{NL'}$:
\be
d/dt(e+\langle \tilde \rho T \rangle /(\gamma-1)) = - Q_{exp} -2\gamma \langle \tilde \rho T \rangle \epsilon/a - Q_{NL'} 
\label{damp2}
\ee
\chb{Fig.~\ref{fig13} shows for run A the total (turbulent + internal) energy time derivative and the associated residual term $Q_{NL'}$}. The residual dissipation is limited to 1 or $2\%$ of the total variation. 
This is substantially smaller than $Q_{NL}$ (see fig.~\ref{fig12}c).
It shows that compressible exchanges between turbulent and internal energy are the dominant contribution to the deviation of turbulent energy conservation during the cascade, especially during the beginning of the evolution.
The same remarks can be made for the other runs in Table~\ref{table1}. However, in run H with a larger expansion parameter ($\epsilon=0.4$), 
we find that compression and expansion contribute more equally (not shown).

\subsection{Conclusion}
In conclusion, using complete nonlinear couplings of MHD equations,
we have shown that radial temperature profiles as $1/R$ simply result from the combination of adiabatic decrease and 
turbulent dissipation.
This has been done, starting at 0.2 AU with an rms Mach number 1 and an expansion parameter $\epsilon=0.2$.
With these parameters, the decrease of rms turbulent amplitude is not much faster than the Alfv\'en wkb prediction, actually as $1/R^{0.6}$.
This demonstration has been done by starting with a spectral anisotropy characteristic of slow winds, namely mainly perpendicular to the mean magnetic field. 
This included showing that the $Q_\nu/Q_{K41}$ ratio is close to 0.1 in our simulations as in the solar wind.
Finally, we measured the deviation from conservation of turbulent energy during the cascade (residual energy loss $Q_{NL}$). We found that expansion was a minor cause of deviation, the main cause being compressible exchanges between
turbulent and internal energy.

Future work includes
(i) providing a clearer signature (i.e., with larger Reynolds) of the $5/3$ power-law index characteristic of slow winds by lowering the Mach number and the expansion parameter; (ii) considering the case of fast winds which\cha{, despite having a different spectral anisotropy,} produce a similar radial dependance of the temperature.

\begin{acknowledgements}
This work was performed using HPC resources from GENCI-IDRIS (grant 2017-040219).
It has been supported by Programme National Soleil-Terre (PNST/INSU/CNRS).
\end{acknowledgements}

\end{document}